\documentclass[journal=jctc,manuscript=article]{achemso}

\usepackage[version=3]{mhchem} 
\usepackage[T1]{fontenc}       
\usepackage{xcolor}  
\newcommand{\todo}[1]{{\color{violet}}}
\usepackage{subscript}
\usepackage{amsmath} 
\usepackage{graphicx}
\usepackage{dcolumn}
\usepackage{bm}
\usepackage[english]{babel}
\usepackage{fancyref}
\usepackage[load=physical]{siunitx}
\usepackage{color}
\usepackage{amssymb}

\def\cm{cm$^{-1}$}
\def\be{\begin{equation}}
\def\ee{\end{equation}}
\def\bea{\begin{eqnarray}}
\def\eea{\end{eqnarray}}

\def\bra#1{\mbox{$\langle#1|$}}
\def\ket#1{\mbox{$|#1\rangle$}}

\author{Per-Arno Pl\"otz}
\affiliation{Institut f\"ur Physik, Universit\"at Rostock,
Albert-Einstein-Str. 23-24, 18059 Rostock, Germany}
\author{J\"org Megow}
\affiliation{Institut f\"ur Chemie, Universit\"at Potsdam, Karl-Liebknecht-Str.
24-25, 14476 Potsdam, Germany}
\author{Thomas Niehaus}
\affiliation{Univ Lyon, Universit\'e Claude Bernard Lyon 1, CNRS,
   Institut Lumi\`ere Mati\`ere, F-69622, Villeurbanne, France}
\author{Oliver K\"uhn}
\email{oliver.kuehn@uni-rostock.de}
\affiliation{Institut f\"ur Physik, Universit\"at Rostock,
Albert-Einstein-Str. 23-24, 18059 Rostock, Germany}

\title{An All-DFTB Approach to the Parametrization of the System-Bath Hamiltonian Describing Exciton-Vibrational Dynamics of Molecular Assemblies}
\begin{document}

%
%
%
%
%
%
%
%
\begin{abstract}
Spectral density functions are central to the simulation of complex many body systems. Their determination requires to make approximations not only to the dynamics but also to the underlying electronic structure theory. Here, blending different methods bears the danger of an inconsistent description. To solve this issue we propose an all-DFTB approach to determine spectral densities for the description of Frenkel excitons in molecular assemblies. The protocol is illustrated for a model of a PTCDI crystal, which involves the calculation of monomeric excitation energies, Coulomb couplings between monomer transitions, as well as their spectral distributions due to thermal fluctuations of the nuclei. Using dynamically defined normal modes, a mapping onto the standard harmonic oscillator spectral  densities is achieved. 
\end{abstract}
%
%
\section{Introduction}
%
The coupled dynamics of electronic and vibrational excitations in molecular assemblies such as aggregates and crystals as well as in pigment-protein complexes is responsible for their optical and transport properties.~\cite{may11} Given the complexity of these systems, one usually resorts to model Hamiltonians, where the Frenkel Hamiltonian with linear exciton-vibrational coupling (Huang-Rhys (HR) model) is the most prominent one.~\cite{renger01_137, schroter15_1} It is typical for so-called system-bath situations. Here, the relevant system is formed by Frenkel excitons (FEs) describing the (collective) electronic excitations, which are coupled to a bath consisting of harmonic vibrations.~\cite{may11}  The system parameters, i.e. the local electronic excitation energies and the Coulomb couplings (CCs) between electronic transitions and charge densities at different monomers,  are readily described using either a fit to experimental data such as spectra or quantum chemical calculations.\cite{krueger98_5378,fuckel08_074505,madjet06_17268} In Ref.~\citenum{plotz14_174101} we have put forward a density functional theory based tight-binding (DFTB) method to efficiently obtain CCs between transitions residing on different monomers. The method is based on the linear response formulation of time-dependent DFTB (TD-DFTB)~\cite{niehaus01_085108} and for the chosen test case showed favorable agreement with results obtained with higher level approaches. The protocol was coined Tight-Binding Frenkel Exciton (TBFE) approach and subsequently applied to the study of the spectroscopic signatures of aggregation behavior of perylene-based aggregates~\cite{ploetz16_25110} as well as to singlet diffusion in crystalline anthracene.~\cite{kranz16_4209}

 The system-bath coupling is usually described by means of a spectral density (SD). Its determination is rather involved as it requires dynamical equations to be solved for the total system. This is usually done by resorting to a quantum-classical approach, i.e. classical molecular dynamics is used for sampling the thermal ensemble in the electronic ground state and electronic transitions and CCs are calculated along these trajectories. This yields respective correlation functions from which spectral distributions of transition energies and CCs can be obtained (e.g. Refs.~\citenum{may11,schroter15_1,arago15_026402,arago16_2316,megow11_645,olbrich11_8609,jing12_1164,shim12_649,fornari16_7987}). The actual SD is defined solely on the distribution of system-bath couplings, i.e.\ independent of temperature. Formally, it follows from a mapping of the real environment onto an effective harmonic bath, which is valid in the linear response limit.~\cite{makri99_2823}

 The method of choice for the calculation of electronic properties and trajectories would be DFT. But, having in mind the potential need  for long trajectories to guarantee sufficient sampling of the thermal fluctuations, it appears to be too costly. Therefore, we have recently extended the TBFE approach to include the effect of thermal motion of the nuclei such as to describe on-site as well as inter-site spectral distributions.\cite{plotz17_084112} In contrast to previous approaches~\cite{megow16_61}, the polarization of the electronic transition densities by the moving nuclei is naturally included within TBFE. Further, using dynamically  defined normal modes~\cite{mathias11_2028}, spectral features in these distributions could be assigned to particular intramolecular motions, thus achieving a rather detailed insight, for instance,  into the origin of the modulation of the CC by specific  vibrations.

In general terms, our efforts are directed towards the development of a simulation protocol, which yields the parameters of the FE Hamiltonian including SDs for a variety of situations, ranging from molecular aggregates in solution or molecular crystal to coupled chromophores in light-harvesting antennae. The key requirement is to use a \textit{single} electronic structure method throughout, i.e. from sampling the thermal ensemble in the electronic ground state to the calculation of electronic excitation energies and CCs. Although this approach will have to sacrifice the accuracy of the electronic structure part, it will be internally consistent and improvable in a controlled manner. The issue of incompatible methods for nuclear dynamics and electronic excitation calculations has previously been coined `geometry mismatch' problem and was addressed, e.g., in Refs. \citenum{rosnik15_5826} and \citenum{lee16_3171} using quantum mechanics/molecular mechanics approaches.

\begin{figure}[bth]
  \centering
  \includegraphics[width=0.6\textwidth]{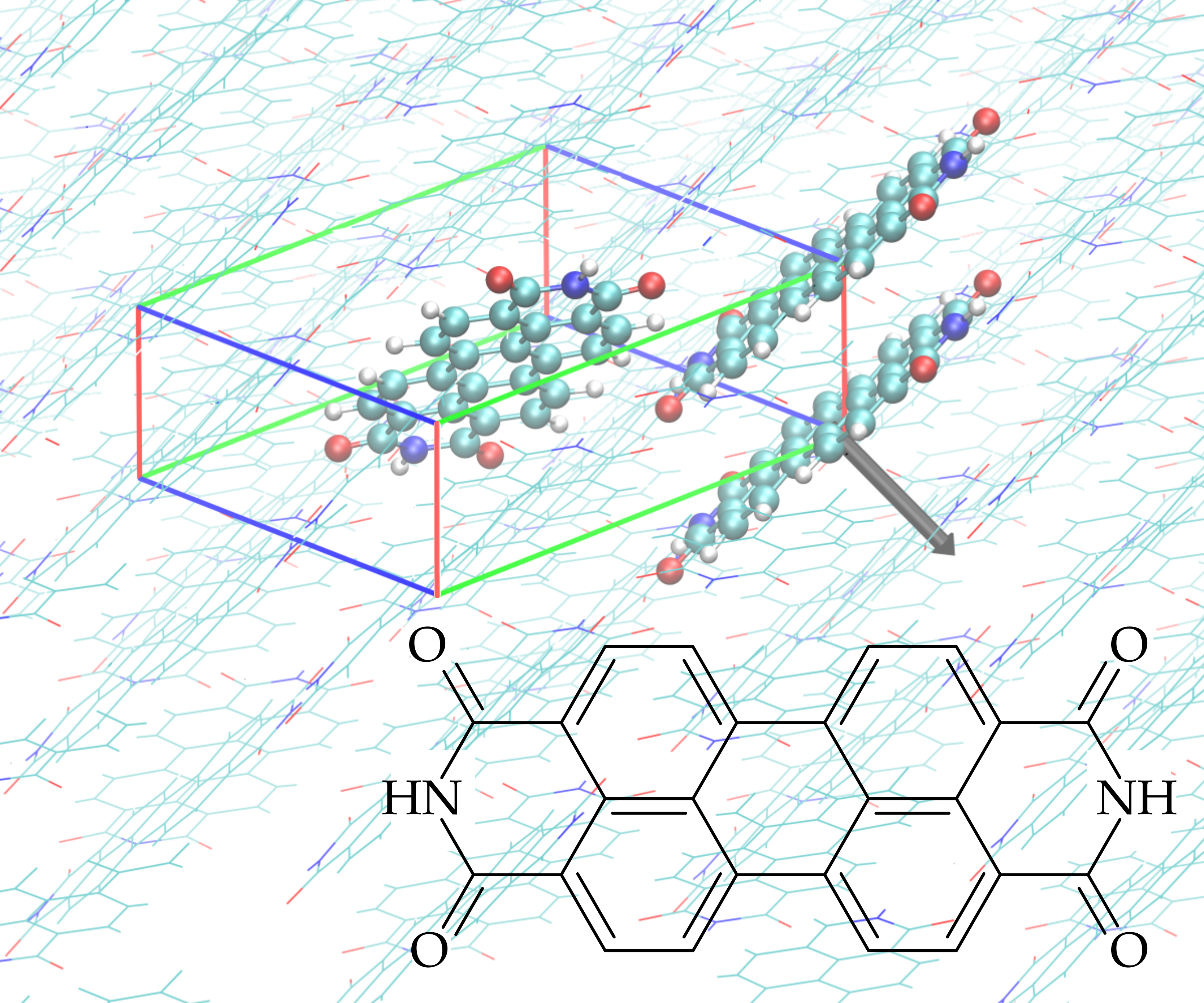}
  \caption{Unit cell of the 3,4,9,10-perylene-tetracarboxylic-diimide (PTCDI) crystal  with its two monomers highlighted (figure generated with VMD\cite{humphrey96_33}); additionally a third monomer is highlighted, which is part of a dimer in stacked configuration (separated by the cell edge). The direction of shifts for distance dependency calculations is depicted as a grey vector. Furthermore the structure formula of  PTCDI is displayed. }
  \label{fig:cell}
\end{figure}

 In Ref.~\citenum{plotz17_084112} we studied a PTCDI (3,4,9,10-perylene-tetracarboxylic-diimide, cf. Fig.~\ref{fig:cell}) crystal using ground state molecular dynamics trajectories, which were based on an empirical force field (FF). This allowed to employ a rather large simulation box containing 216 PTCDI molecules. Analyzing the obtained spectral distributions in terms of normal mode frequencies we have noticed that an approximate agreement between FF and DFTB frequencies can be achieved by uniform scaling of the former to match a peak around 1550 \cm{} dominating the on-site spectral distribution. The purpose of the present contribution is to exploit the full potential of the TBFE approach by presenting an all-DFTB analysis of the FE Hamiltonian including effective SDs. As in Ref.~\citenum{plotz17_084112} we will use the exemplary case of  a PTCDI crystal. As a side effect, this will also put us into the position of assessing the issue of a `not consistent' simulation that combines empirical forces for the trajectory with excitation energies and densities derived from electronic structure theory. 

For the PTCDI crystal, like for many other systems with closely packed chromophores, the relevance of intermolecular charge transfer (CT) states is frequently discussed.\cite{gisslen09_115309,yamagata11_204703,megow15_5747,forker15_18816,megow15_18818,settels12_1544,bellinger17_2434} Of course, TD-DFTB inherits the problems with CT states from DFT, i.e. care must be taken to address the CT issue. Concerning the intramolecular CT-like effects, it had been shown in Ref.~\citenum{plotz17_084112} that for the lowest bright transition TD-DFTB yields the correct fluctuation dynamics, although the average energy is slightly shifted, if compared with TD-DFT using the long-range corrected  CAM-B3LYP functional. The more important intermolecular CT cannot be handled with the standard FE Hamiltonian. To explore their potential effect on the energetics of PTCDI crystals, in the present contribution we have systematically investigated a stacked PTCDI dimer using different DFT functionals as well as a wavefunction-based supermolecule calculation in conjunction with a simple four level model Hamiltonian.

The paper is organized as follows: In Section~2 the derivation of the TBFE Hamiltonian, the characterization of CT states, and the introduction of SDs are briefly sketched. Computational details are summarized in Section~3. The results section~4 starts with an investigation of the role of CT states. Next, spectral distributions of TBFE parameters will be discussed along with the mapping onto an effective harmonic oscillator bath. Finally, a brief account on the electronic  absorption spectrum is given. Summary and conclusions are presented in Section~5.
%

%
\section{Theory}
\label{sec:theory}
\subsection{TBFE Hamiltonian Within Linear Response TD-DFTB}
\label{ssec:dftb}

In the following we will briefly outline the adaption of the Casida approach~\cite{casida95_155} to obtain transition energies and densities within linear response TD-DFTB (for details, see Ref.~\citenum{niehaus01_085108}). The Casida approach is based on the eigenvalue equation \cite{casida95_155,stratmann98_8218}
\begin{equation}
\label{eq:casida1}
\begin{pmatrix}
\bf{A}&{\bf B} \\
{\bf B} & {\bf A} \\
\end{pmatrix}
\begin{pmatrix}
{\bf X}\\
{\bf Y}\\
\end{pmatrix}
=\omega\begin{pmatrix}
1&0 \\
0&-1 \\
\end{pmatrix}\begin{pmatrix}
{\bf X}\\
{\bf Y}\\
\end{pmatrix}\,.
\end{equation}
The eigenvalues and eigenvectors are the transition energies and densities, respectively.
In the above equation we have used
\begin{equation}
\label{eq:A}
A_{is,jt}=\delta_{ij}\delta_{st}\omega_{is}+K_{is,jt}
\end{equation}
and 
\begin{equation}
\label{eq:B}
B_{is,jt}=K_{is,tj}\,.
\end{equation}
Here, we focus on closed-shell systems and singlet excitations
only. The term $\omega_{is}=\epsilon_s-\epsilon_i$ stands for the
energy difference between an occupied Kohn-Sham orbital ($i,j,\ldots$)
and a virtual one ($s,t,\ldots$). Within TD-DFTB the coupling matrix
$K_{is,jt}$ is approximated as\cite{niehaus01_085108,kranz17_1737} 
\begin{equation}
\label{kdftb}
K_{is,jt}=\sum_ {AB}\left(2q^{is}_A\gamma_{AB}^\text{fr}q^{jt}_B - c_x q^{ij}_A\gamma_{AB}^\text{lr}q^{st}_B\right)
\end{equation}
where the sum is with respect to the atoms
$(A,B,\ldots)$. Eq.~(\ref{kdftb}) summarizes conventional
DFTB ($c_x=0$) with local/semi-local exchange-correlation functionals and the recent extension LC-DFTB ($c_x=1$) which
involves a contribution of Hartree-Fock exchange in a range-separation
formalism.~\cite{niehaus12_237,lutsker15_184107,kranz17_1737} The terms $\gamma^\text{fr/lr}_{AB}$ are approximated two-electron integrals giving a measure for the strength of the electron-electron interaction. Further, we have introduced the one-electron Mulliken transition charges
\begin{equation}
	q^{is}_A= \frac{1}{2}\sum_{\mu\in{}A}\sum_{\nu}(C_{\mu i} C_{\nu s} S_{\mu\nu}+C_{\nu i} C_{\mu s} S_{\nu\mu})\,.
\end{equation}
Here, the $C_{\mu i}$ are the molecular orbital (MO) coefficients resulting from the expansion of the one-electron MOs into atomic basis functions, i.e. $\psi_i({\bf r})=\sum_\mu C_{\mu i} \phi_\mu({\bf r})$; the respective matrix will be denoted as ${\bf C}$. Further, the  $S_{\mu\nu}$ are the  elements of the overlap matrix, ${\bf S}$, in the atomic orbital basis.

Solving Eq.~(\ref{eq:casida1}) one obtains the transition energies
$\omega_I$ for
a given excited singlet state $I$, as well as the associated many-electron transition densities
\begin{equation}
\label{rhotrans}
	\rho_I({\bf r}) = \sqrt{2}\sum_{is} \left(\bf{X}+\bf{Y}\right)^{I}_{is} \psi_i({\bf r})\psi_s({\bf r})\,.
\end{equation}
Using the approximation $\psi_i({\bf r})\psi_s({\bf r})\approx \sum_A q_A^{is}  \Phi_A({\bf r})$, with $\Phi_A({\bf r})=N_A^{-1}\sum_{\mu \in A} |\phi_\mu({\bf r})|^2 $ ($N_A$ is the number of basis functions on atom $A$), this can also be written as
\begin{equation}
	\rho_I({\bf r}) = \sum_A Q_A^I  \Phi_A({\bf r})\, .
\end{equation}
Here, we introduced the atom-centered many-electron Mulliken
transition charges according to
\begin{equation}
	Q_A^I =  \sqrt{2}  \sum_{is}\left(\bf{X}+\bf{Y}\right)^{I}_{is} q_A^{is} \,.
\end{equation}

In the following we will specify the situation to an assembly of two-level chromophores, resonantly coupled via their transition densities and neglecting exchange contributions. The ground and excited state of monomer $m$ is denoted as $|g_m\rangle$ and $|e_m\rangle$, respectively. Restricting ourselves to one-exciton states, $|m\rangle = |e_m\rangle \prod_{k\ne m} |g_k\rangle$, only, the FE Hamiltonian reads~\cite{may11}
\begin{equation}
\label{eq:Ham_hmn}
H_{\rm ex}=\sum_{mn}(\delta_{mn}E_m+J_{mn})\ket{m}\bra{n}=\sum_{mn}h_{mn}\ket{m}\bra{n}
\end{equation}
with the local excitation energy, $E_m$, and the CC $J_{mn}$. Within the TBFE approach the latter is expressed as (note the identification $I\equiv eg$ where the monomer index follows from the summation of atomic labels)~\cite{plotz14_174101}
\begin{equation}\label{eq:trans_trans}
J_{mn}=\sum_{A\in m}\sum_{B\in n}Q^{eg}_A Q^{ge}_B\zeta_{AB}(|{\bf R}_A-{\bf R}_B|) \, .
\end{equation}
Here, $\zeta_{AB}$ is a Coulomb integral taken with respect to the functions $\Phi_A$ and $\Phi_B$, which are centered at ${\bf  R}_A$ and ${\bf  R}_B$, respectively.
%
\subsection{Spectral Densities and Normal Modes}
%
The one-exciton Hamiltonian matrix, $h_{mn}$, depends on the nuclear coordinates. Hence, inter- and intramolecular nuclear motions characterizing the thermal ensemble render the Hamiltonian matrix time-dependent, i.e. $h_{mn}(t)$. Upon introducing averaged energies, $\langle h_{mm}\rangle$, and the respective fluctuations, $\delta h_{mn}(t)=h_{mn}(t)-\langle h_{mm}\rangle$, Eq. (\ref{eq:Ham_hmn}) becomes
\begin{equation}
H_{\rm ex}=\sum_{mn}\left[\langle h_{mm}\rangle+\delta h_{mn}(t)\right]\ket{m}\bra{n}\, .
\end{equation}
The fluctuations can be characterized by means of their correlation functions and the related spectral distributions, i.e.
\begin{equation}
\label{eq:spec_dens}
C_{kl,mn}(\omega)=\int dt e^{i\omega t}\langle \delta h_{kl}(t)\delta h_{mn}(0)\rangle\,.
\end{equation}
In the early days of exciton theory,  explicit calculation of the fluctuation correlation function was out of reach and, therefore, models such as the Haken-Strobl-Reineker~\cite{haken73_135,reineker82_111} parametrization have been very popular. It  assumed Gaussian-Markovian statistics, leading to a high-temperature stochastic Liouville equation. Later, \v{C}apek and coworkers generalized the Haken-Strobl model to include finite temperature situations.~\cite{capek85_667}

In the following our goal is to provide a means for interpretation of $C_{kl,mn}(\omega)$. Since nuclear motions are typically of small amplitude, a harmonic oscillator bath model is appropriate.~\cite{kuhn97_213,may11,renger01_137,schroter15_1} To lowest order in the (dimensionless) normal mode coordinates, $Q_\xi$, the fluctuations can be written as
\begin{equation}
	\delta h_{mn}(t)= \sum_\xi \hbar \omega_\xi g_{mn}(\xi) Q_\xi(t) \, ,
\end{equation}
where $g_{mn}(\xi)$ is a (dimensionless) coupling matrix. In fact, in Ref.~\citenum{plotz17_084112} it had been shown that the approximation  of linear coupling is well justified for the present system. This approximation enables one to evaluate Eq.~(\ref{eq:spec_dens}) analytically yielding~\cite{may11} 
\begin{equation}
\label{eq:spec_dens_HO}
C^{\rm osc}_{kl,mn}(\omega)= 2\pi \omega^2 [1+n(\omega)][{\mathcal J}_{kl,mn}(\omega) - {\mathcal J}_{kl,mn}(-\omega)  ] \,. 
\end{equation}
Here, $n(\omega)$ represents the Bose-Einstein distribution and the harmonic oscillator SD has been introduced according to
\begin{equation}
{\mathcal J}_{kl,mn}(\omega)=\sum_\xi g_{kl}(\xi)g_{mn}(\xi)\delta(\omega - \omega_\xi)\, .
\label{eq:SD} 
\end{equation}
In general, the total set of modes can be partitioned into intra- and intermolecular (phonon) modes. Since the present simulation box is too small to reliably discuss phonon modes, we will restrict the following discussion to intramolecular modes only. For simplicity we assume that each molecule has its own set of intramolecular modes, whose characteristics, however, are the same   for all molecules.  Note that for the inter-site SDs, the index $\xi$ runs over modes from all involved sites. Following Ref.~\citenum{plotz17_084112} we will use generalized (dynamically defined) normal modes obtained by minimizing cross-correlations within a tensorial extension of the vibrational density of states.~\cite{mathias11_2028} This takes into account  anharmonicity of the system as the density of states is obtained on the basis of a molecular dynamics trajectory. In other words, the actual anharmonic dynamics is mapped onto that of a harmonic oscillator bath.

Having at hand the generalized normal modes, the coupling matrix can be obtained via
\begin{equation}
\label{eq:gmn}
	g_{mn}(\xi) = \frac{1}{\hbar \omega_\xi}\left.\frac{\partial h_{mn}}{\partial Q_\xi}\right|_{Q_\xi=0} \,.
\end{equation}
In practice this expression is obtained by a finite difference approximation starting from a structure, averaged with respect to the molecular dynamics trajectory. 
%
\subsection{Measure of Charge Transfer Character for TD-DFTB}
\label{ssec:CT_DFTB}
There are  algorithms for the calculations of charge transfer characteristics of excitations in supermolecular DFT and coupled cluster approaches.~\cite{theodore15} However, here  we present a respective formulation for TD-DFTB calculations.  Following  the concept by Plasser and Lischka~\cite{plasser12_2777} the charge transfer share is calculated by means of the charge transfer numbers 
\begin{equation}
\Omega^{I}_{mn}=\frac{1}{2}\sum_{\mu\in{}m,\nu\in{}n}\left(\mathbf{D}^{I}\mathbf{S}\right)_{\mu\nu}\left(\mathbf{S}\mathbf{D}^{I}\right)_{\mu\nu}
\end{equation}
for the monomers $m$ and $n$ (with atomic orbitals $\mathrm{\mu}$ and $\mathrm{\nu}$ and excitation $I$).  Besides the atomic orbital overlap matrix, ${\bf S}$, one needs the one-electron transition density matrix, $\mathbf{D}^{I}$, in atomic orbital basis. It follows from the  MO coefficients and the eigenvectors $\mathbf{X}^{I}$ and $\mathbf{Y}^{I}$ of the Casida equation~\cite{marques06_,furche02_7433,casida95_155}
\begin{equation}
\mathbf{D}^{I}=\sqrt{2}\,\mathbf{C}\left(\mathbf{X+Y}\right)^I\mathbf{C}^\mathrm{T}\,.
\end{equation}

The so-called charge transfer number can be calculated as
\begin{equation}
\label{eq:CT}
{\rm CT}_I=\frac{1}{\Omega^{I}}\sum_{m, n\neq{m}}\Omega^{I}_\mathrm{mn}
\end{equation}
with the sum  $\Omega^{I}=\sum_{m,n}\Omega^{I}_{mn}$. Since all ingredients in Eq.~\eqref{eq:CT} are readily available during a TD-DFTB calculation the CT character can be evaluated with comparatively small effort.
%
\subsection{Four Level Hamiltonian}
\label{sec:4LS}
%
Below, the issue of mixing between FE and CT states will be addressed using a dimer supermolecule model.  In order to map the adiabatic electronic states of this supermolecule model onto a local diabatic model, we will use a simple four level Hamiltonian (for a more elaborate treatment of systems having arbitrary size, see Ref. ~\citenum{engels17_12604}). The pair of FE states will be described by the Hamiltonian given in Eq.~\eqref{eq:Ham_hmn} ($E_1=E_2=E_{\rm FE}$ and $J_{12}=J$), with the eigenstates $|{\rm FE}_+\rangle$ and $|{\rm FE}_-\rangle$ having energies $E_\pm=E_{\rm FE} \pm J$. It has to be supplemented by a pair of CT states, where the linear combinations (eigenstates) are called $|{\rm CT}_+\rangle$ and $|{\rm CT}_-\rangle$. They are assumed to be degenerate with energy $E_{\rm CT}$, i.e. their coupling leading to a simultaneous transfer of both charge pairs is neglected. Finally, we introduce a mixing between FE and CT states, which leads to the four-level Hamiltonian matrix in the basis $\{$\ket{\rm{FE}_+}$, $\ket{\rm{CT}_+}$,  $\ket{\rm{FE}_-}$, $\ket{\rm{CT}_-}$ \}$
\begin{equation}
\label{eq:4LS}
	H^{\rm 4L}=\left(\begin{matrix}
E_{\rm FE}+J & V_+ & 0 & 0 \\
 V_+ & E_{\rm{CT}} & 0 & 0 & & &\\
  0 & 0 & E_{\rm FE}-J & V_-\\
  0 & 0 & V_- &  E_{\rm{CT}}
\end{matrix}\right) \, .
\end{equation}
Note that only states of the same gerade/ungerade symmetry interact with strength denoted as  $V_\pm$.


This Hamiltonian is readily diagonalized, yielding the energies

\begin{eqnarray}
\label{eq:energies}
E^{u/g}_1&=&\frac{1}{2}\left[E_\pm +E_{\rm CT}-\left((E_\pm -E_{\rm CT})^2+4V^2_\pm\right)^{1/2}\right]\nonumber\\
E^{u/g}_2&=&\frac{1}{2}\left[E_\pm +E_{\rm CT}+\left((E_\pm -E_{\rm CT})^2+4V^2_\pm\right)^{1/2}\right] \,.\nonumber\\
\end{eqnarray}

In principle all parameters depend on the geometry of the two monomers within the dimer, i.e. the distance and relative orientation. Below the latter will be fixed and only the intermonomer distance will be varied to describe adiabatic and diabatic potential energy curves.

\section{Computational Methods}
\label{sec:cmeth}
All-DFTB simulations have been performed to parametrize the TBFE Hamiltonian for  a PTCDI crystal. For reasons of numerical feasibility, the simulation box has been  reduced to 2$\times$2$\times$2 primitive cells each containing two molecules as shown in Fig.~\ref{fig:cell} (i.e. there is a total of 16 molecules as compared to 216 in Ref.~\citenum{plotz17_084112}). In contrast to Ref. \citenum{plotz14_174101}, where the crystalline structure was approximated by an orthorhombic cell to be comparable to Ref. \citenum{megow15_5747}, the present work uses the triclinic cell as described in Ref.~\citenum{klebe89_69}.

DFTB ground state trajectory simulations have been performed using the DFTB+ software\cite{aradi07_5678} and the mio-1-1 Slater-Koster parameter set.~\cite{elstner98_7260} In order to quantify the influence of the used method for the MD forces, an additional trajectory is propagated employing the AMBER force field \cite{case04} and the GAFF parameter set \cite{wang04_1157} and using the  NAMD program package.~\cite{phillips05_1781} Both simulations are performed with a time step of 1 fs and at a temperature of 300~K in a canonical ensemble. After  equilibration runs of 1~ns (FF) and 100 ps (DFTB), sampling was carried out during production runs of 100~ps.  

Based on these trajectories,  the TBFE parameter and correlation function calculations were done  according to the protocol described in Ref.~\citenum{plotz17_084112}. To aid analysis of the spectral distributions and to obtain a mapping to harmonic oscillator SDs we have determined generalized normal modes for both trajectories using the method of Mathias and Baer~\cite{mathias11_2028}. The thus defined normal mode displacements are used to obtain the exciton-vibrational coupling parameters according to Eq.~\eqref{eq:gmn}. HR factors follow as $S_m(\xi)=[g_{mm}(\xi)]^2/2$, Further, absorption spectra are sampled from the trajectory data. To this end the TBFE Hamiltonian is diagonalized at each time step using periodic boundary conditions. Here, only the strongest CCs are taken into account, i.e. for the stacked, head-to-tail and step motifs. Absorption spectra are calculated from the matrix elements of the molecular transition dipole operators, which are taken according to the average geometry.

The used unit cell is depicted in Fig. \ref{fig:cell}, together with an exemplary pair of monomers in the stacked configuration.  For the discussion of CT mixing, we concentrate on the stacked configuration having the strongest CC as compared to other motifs (cf. Ref.~\citenum{plotz17_084112}). Further, it is expected to be the configuration which is influenced the most by CT states due to the overlap of the monomer charge densities, which is obviously smaller for every other relative configuration.~\cite{bredas04_4971}

The mixing between Frenkel and CT transitions will be characterized using a dimer model. First, the PTCDI monomer geometry in gas phase is determined at D$_{\rm 2h}$ symmetry using the DFT/CAM-B3LYP/def2-TZVP setup as implemented in Gaussian 09.~\cite{g09} The dimer structure is generated by a least-square optimization of two of these optimized monomers to the average atomic positions of the stacked configuration along the trajectory (cf. Fig.~\ref{fig:cell}). Subsequently, the monomers are displaced in direction of the vector, normal to the monomer plane as shown in Fig.~\ref{fig:cell}. The distance $R$ is counted with respect to the averaged value which is $4.8~$\AA{} defined as $R=0~$\AA.

For a representative number of distances, electronic supermolecule calculations are performed with five different methods. As a reference we will use a SCS-CC2 calculation (SVP basis set, TURBOMOLE implementation~\cite{turbomole6.5}). Further, we included long-range corrected TD-DFT/$\omega$B97X/def2-TZVP and TD-DFT/CAM-B3LYP/def2-TZVP as implemented in Gaussian09.~\cite{g09} Finally, concerning the TD-DFTB methods, we have used standard DFTB (mio-1-1) as well as LC-DFTB (modified mio-1-1 parameters\cite{kranz17_1737}, range-separation parameter $\omega$ is set to $\omega= $ 0.3 a$_{0}^{-1}$).

The thus obtained distance dependent energetics will be fit to the four-level model introduced above. Here, $E_{\rm FE}$  follows from the transition energy of the FE states at large separations. The other four parameters can be obtained by rearranging Eqs.~\eqref{eq:energies}.

It is known that the mixing between FE and CT states can be influenced by environmental charges. To quantify this effect we have repeated the supermolecule calculations for $R=0$~\AA, but now including the immediate environment (first shell) as classical point charges. Here, the DFTB Mulliken charges of a gas phase monomer are used for simplicity. They are placed at the average atomic positions taken from the DFTB trajectory. The dimer (at the $R=0$~\AA) geometry is placed, relative to the environment, by minimizing the sum of distances to the respective averaged atomic positions.

Finally, for all structures and methods  CT numbers, ${\rm CT}_I$, are calculated. In case of  DFT and SCS-CC2 the Theodore program is used.~\cite{theodore15,plasser12_2777} For the TD-DFTB methods the calculation of ${\rm CT}_I$ has been implemented as described in  Section~2. 

\begin{figure*}[th]
  \centering
  \includegraphics[width=0.95\textwidth]{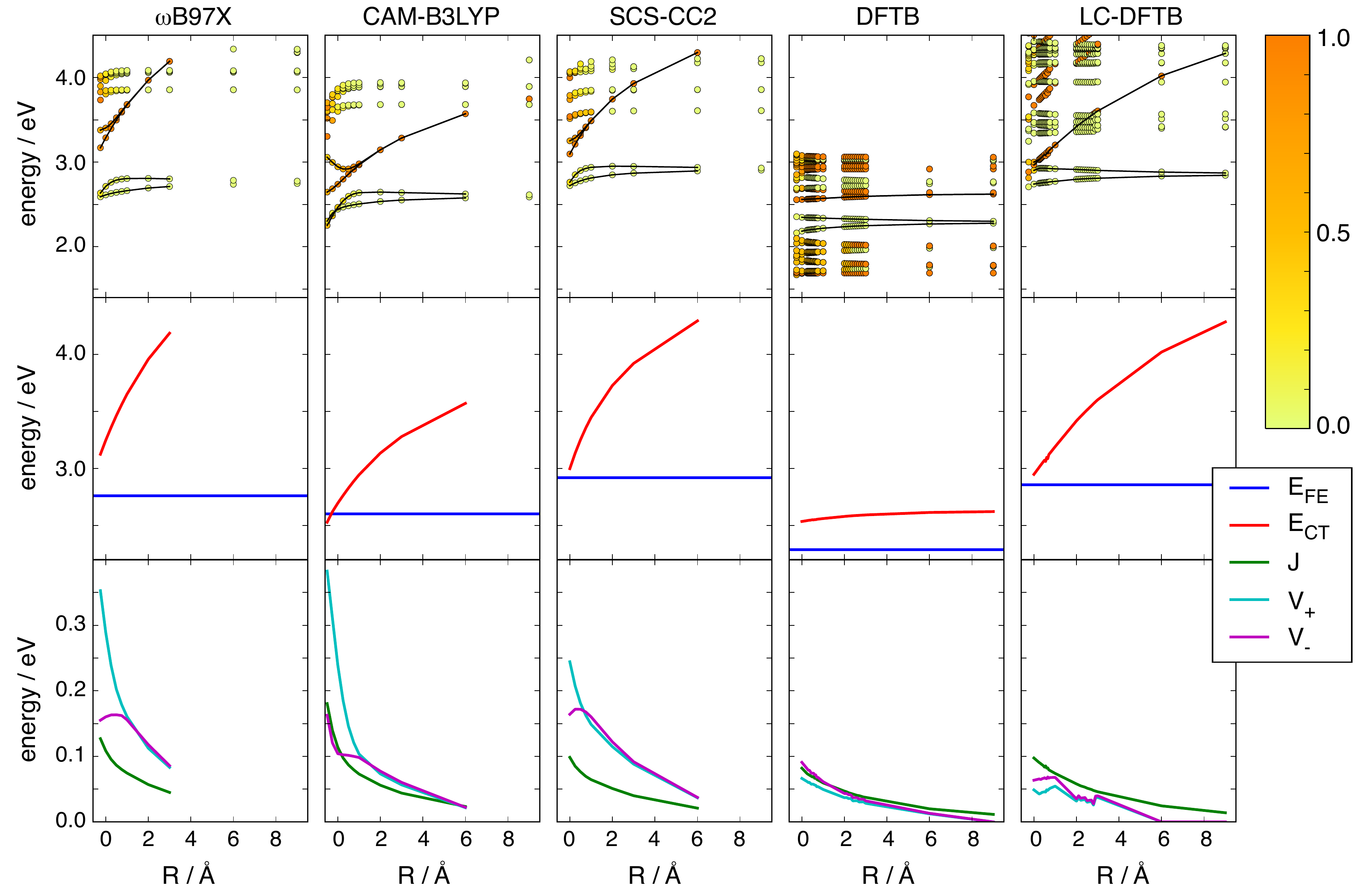}
  \caption{Charge transfer characters, CT$_I$, of an isolated PTCDI dimer (taken from a crystalline structure) depending on intermolecular distance ($R=0$~\AA{} corresponds to the crystal structure, see text for details). The calculations were performed employing different electronic structure methods as indicated. The top row shows excitation energies and CT$_I$ values (color code given at the right). The solid black lines result from a fit of the excitation energies to Eq.~\eqref{eq:energies}. The two lower rows give the respective parameters of the model Hamiltonian, Eq.~\eqref{eq:4LS}.}
  \label{fig:ct_5_methods}
\end{figure*}
%
\section{Results}
\label{sec:Results}
\subsection{Quantification of CT Mixing}
The dependences of the low-lying excitations on the intermolecular distances of a stacked dimer are shown in Fig.~\ref{fig:ct_5_methods} for different methods. The color code gives the CT character according to Eq.~\eqref{eq:CT}. First, we notice that the results obtained for SCS-CC2 and both DFT functionals (CAM-B3LYP and $\omega$B97X) are qualitatively rather similar. Here, the two lowest transitions are dominantly of local FE character. For distances exceeding 2~\AA{} they show a behavior, which is typical for FE states, i.e.\ a symmetric splitting with respect to the monomer transition energy which decreases with increasing distance. For $R<2$~\AA{} one observes a mixing with a pair of CT states, which results in a total of four transitions, with the upper two retaining their substantial CT character. DFT and SCS-CC2 predict quantitative differences with respect to the strength of the mixing between CT and FE states. Taking SCS-CC2 as reference, $\omega$B97X performs better than CAM-B3LYP.

Figure~\ref{fig:ct_5_methods} also contains the results for time-dependent DFTB and LC-DFTB. DFTB provides no reasonable description of CT states. In fact, the $1/R$-like dependence of CT state energies is not seen at all and there is even a number of CT states below the FE states. Both types of states do not mix noticeable with each other. In contrast, LC-DFTB yields a reasonable distance dependence  for the CT states. Compared to SCS-CC2, however, the mixing between both types of states sets in at a shorter distance. 

The behavior of the coupled CT and FE states is rather well described
in terms of the four state model, Eq.~\eqref{eq:4LS} (solid lines in Fig.~\ref{fig:ct_5_methods}, top row). The middle and the bottom row of Fig.~\ref{fig:ct_5_methods} show the results of the decomposition of the adiabatic dimer energies in terms of the diabatic state parameters of Eq.~\eqref{eq:4LS}. Overall, $\omega$B97X and SCS-CC2 show a rather good agreement. As far as the energies $E_{\rm FE}$ and $E_{\rm CT}$ as well as the FE coupling $J$ are concerned, also LC-DFTB performs well. The coupling between CT and FE states, $V_\pm$, does not seem to be well-described by LC-DFTB. Note that DFTB underestimates $E_{\rm FE}$ (cf. Ref.~\cite{stojanovic17_5846}), whereas the CC, $J$, is  reasonably reproduced.

CT character and excitation energies at the equilibrium configuration are not appreciably modified if fixed environmental point charges are taken into account (not shown).  In fact, we expect that both CT and FE state energies depend strongly on interactions with the environment. It was shown in Ref. \citenum{megow16_94109} that dispersive (i.e. induced-dipole--induced-dipole) interaction strongly shifts FE states. We believe that for CT states inductive (i.e. dipole--induced-dipole) interaction may be of even larger importance. These effects are not included in a description of the intermolecular interaction via fixed point charges and their inclusion will be inevitable in order to finally quantify CT mixing.

\subsection{Spectral Densities}
\begin{figure}[th]
  \centering
  \includegraphics[width=0.9\textwidth]{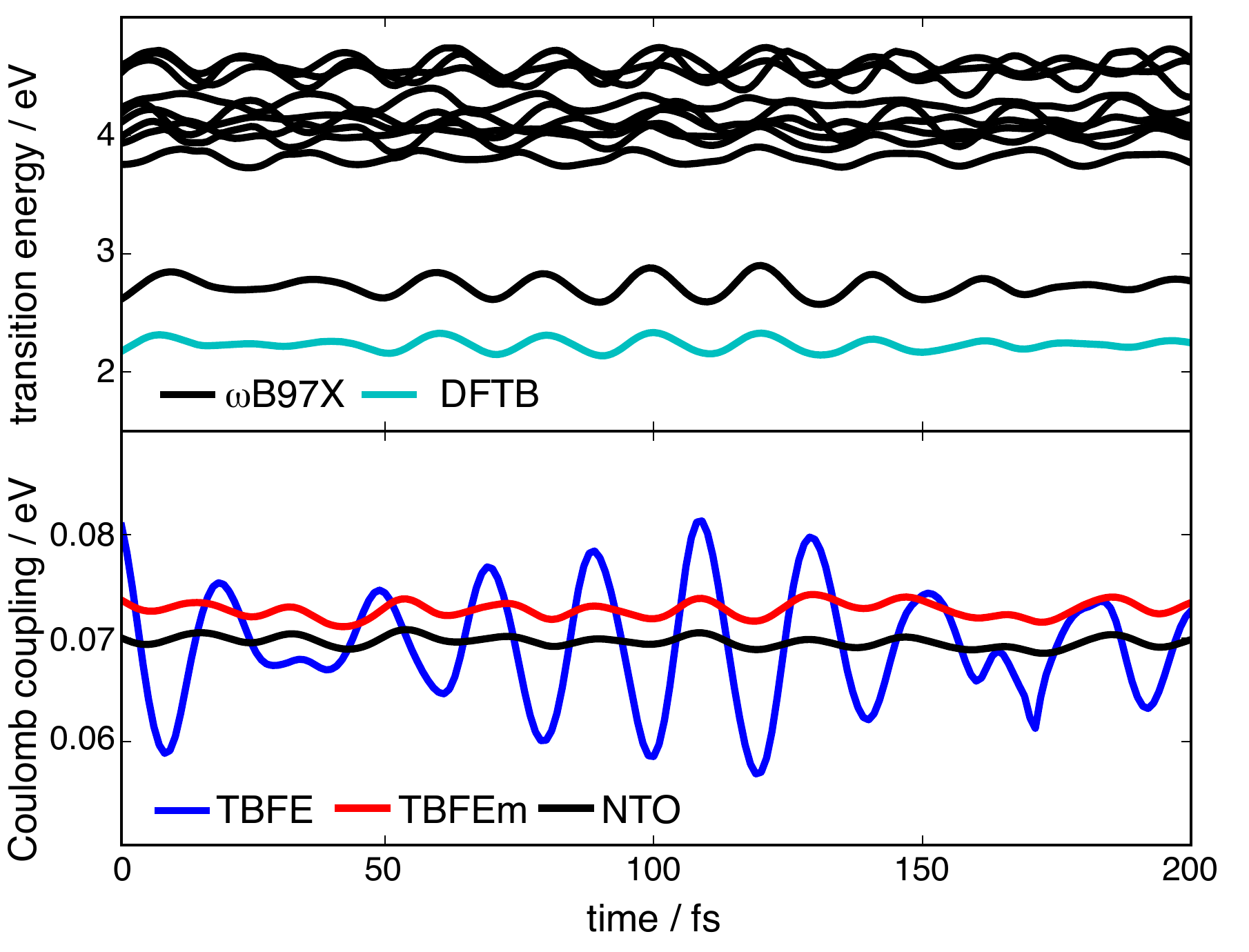}
  \caption{Upper panel: Thermal fluctuations of TD-DFT ($\omega$B97X/TZVP, black) and DFTB (green, first excitation with non vanishing oscillator strength) transition frequencies along a part of the DFTB trajectory. Lower panel: Fluctuations of the  CC between $S_0-S_1$ transitions of monomers in a stacked configuration (one of them considered in the  upper panel). Couplings are compared for the cases of  TBFE, TBFEm (averaged Mulliken transition charges), and NTO (fixed natural transition orbitals); see Ref. \citenum{plotz17_084112} for further details).}
  \label{fig:energies_temporal}
\end{figure}

%
The analysis of the previous section suggests that TBFE parameters can be extracted from  DFTB calculations, provided that the local excitation energy is adjusted.  In Fig.~\ref{fig:energies_temporal} (upper panel) we show a comparison between DFTB and DFT/$\omega$B97X electronic excitation energies along a DFTB trajectory. In case of $\omega$B97X the lowest bright transition is well separated from a band of dark transitions. DFTB predicts a number of low-lying dark transitions with intramolecular CT character (not shown). The first bright transition is slightly below the one obtained using DFT/$\omega$B97X (cf. Fig. \ref{fig:ct_5_methods}). Most importantly, however, the temporal dependence of the DFTB and DFT/$\omega$B97X bright state energies matches very well. This justifies the use of the DFTB method for characterizing the on-site SDs in the following.

The lower panel of Fig.~\ref{fig:energies_temporal} shows the CC for a stacked dimer along the same trajectory. Here, the TBFE approach is compared to the case where the Mulliken transition charges are fixed at their average values (TBFEm). As already pointed out in Ref.~\citenum{plotz17_084112}, TBFE captures the effect of polarization of the transition charges due to intramolecular vibrational motion. In case of fixed transition charges this effect is missing. Here, the small amplitude oscillations indicate that only the interaction of the  fixed charges is modulated by vibrations. In passing we note that in case of the much larger box used in Ref.~\citenum{plotz17_084112} there have been additional slow modulation components due to phonon type intermolecular motions. Figure~\ref{fig:energies_temporal} also contains a comparison with the case of interacting transition charges obtained using fixed NTO (natural transition orbitals) analysis of the monomeric CAM-B3LYP transition density. The NTO result shows a similar oscillation pattern as the TBFEm one, although the average values of the CC are slightly different. This comparison can be taken as an indication that TBFE will provide inter-site SDs which are of similar quality as those obtained using a DFT method. 

Finally, we comment on the use of bare CCs. In Ref.~\citenum{arago15_026402} it had been shown for anthracene crystals that due to the close packing charge density overlap and exchange effects do influence the coupling between adjacent monomers. For the present case we have estimated such effects by comparing the splitting of FE states in a stacked dimer as obtained from a supermolecule calculation (DFTB in Fig.~\ref{fig:ct_5_methods}) or using the TBFE approach. Taking the trajectory-averaged geometry of the dimer one obtains a splitting of 0.152~eV and 0.148~eV in the former and latter case, respectively. The good agreement justifies the use of the bare CCs for the present system.

\begin{figure}[th]
  \centering
  \includegraphics[width=0.90\textwidth]{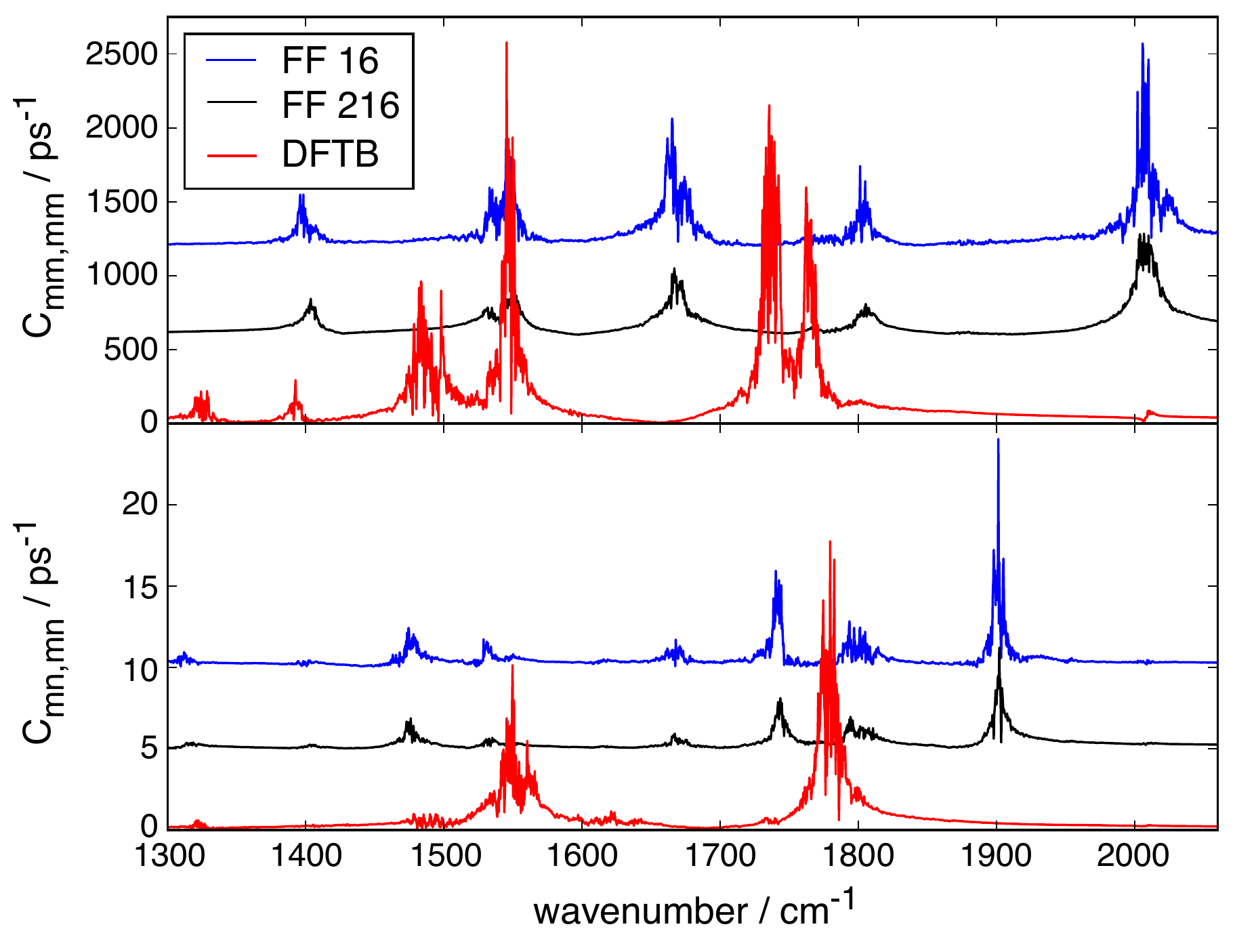}
  \caption{Spectral distribution of fluctuations of the on-site energies, $C_{mm,mm}$, (upper panel) and the CC, $C_{mn,mn}$, (lower panel, stacked motif).  The different curves correspond to the DFTB and the FF trajectories including 16 monomers and the FF trajectory for the case of 216 monomers. }
  \label{fig:energies_traje}
\end{figure}
\begin{figure}[th]
  \centering
  \includegraphics[width=0.9\textwidth]{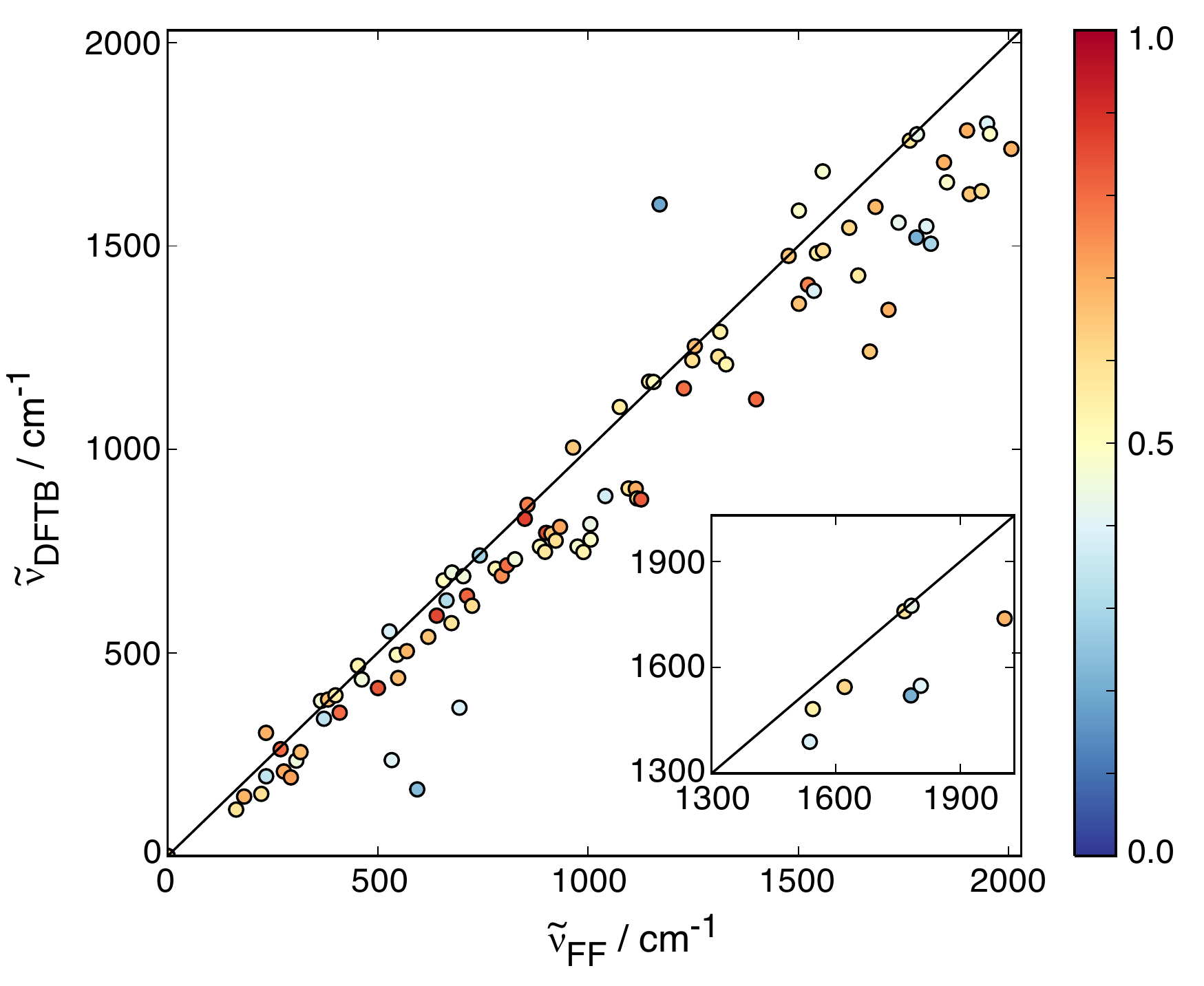}
  \caption{Correlation between effective normal modes from DFTB  and FF trajectories (16 monomers).  The assignment of the modes is made by maximizing the total overlap according to the Kuhn-Munkres algorithm.~\cite{munkres57_32} The inset show those modes having a HR factor greater than 0.02 in the frequency  range given in Fig. \ref{fig:energies_traje}. Note that when using the overlap as the only criterion, it is likely to obtain a few outliers (compare Tab.~1). They have been omitted in this plot by showing assignments with $|\tilde{\nu}_{\rm{FF}}-\tilde{\nu}_{\rm{DFTB}}|<500$~\cm{} only.}
  \label{fig:overlap}
\end{figure}

Figure \ref{fig:energies_traje} shows a comparison of spectral distributions obtained for on-site and CCs using different forces and box sizes. Due to the small box size in the present simulation we cannot reproduce the low frequency phonon-type intermolecular motions discussed in Ref.~\citenum{plotz17_084112}.  Therefore, the following discussion is restricted to the range of intramolecular vibrations (C-C stretch etc.) from 1300 to about 2000 \cm. In passing we note that this is the range where the present TBFE method is superior to approaches using fixed transition densities; cf. Fig.~\ref{fig:energies_temporal}.  First, we notice that the positions of the peaks in the two FF simulations are essentially unchanged, what is to be expected from their intramolecular nature. Since the correlation functions have been averaged with respect to the monomers or pairs in the simulation box, the results look smoother in case of the large box. Comparing DFTB and FF trajectory results for the small box we notice a substantial difference. For instance, FF predicts the most intense peak for $C_{mm,mm}(\omega)$ at about 2000~\cm, whereas with DFTB one obtains about 1750~\cm. 

It is instructive to make a more quantitative comparison between FF and DFTB vibrational features. To this end effective normal modes have been determined for both trajectories. Their correlation based on the overlap of displacement vectors is shown in Fig.~\ref{fig:overlap} according to the Kuhn-Munkres algorithm.~\cite{munkres57_32} In general, the FF frequencies are systematically larger than the DFTB ones. This might suggest that the shortcomings of the FF could be overcome by introducing a scaling as proposed in Ref.~\citenum{plotz17_084112}. Taking the DFTB modes as a reference, however, the agreement thus obtained would be doubtful if quantitative interpretations, e.g., based on the nature of the modes are required. This can be seen from the values of the overlaps in Fig.~\ref{fig:overlap}. Although  the overlap between modes obtained with the two methods is above about 50\% in most cases, only a for a few modes  80\% is exceeded. Of course, this might be acceptable if only peak positions and intensities are required, e.g. in quantum dynamics simulations, see e.g. Ref.~\citenum{schulze16_185101}. However, due to the small overlap, Huang-Rhys factors will also differ as can be seen in Tab.~1, and thus the spectral distributions have to be used with care in a quantitative analysis. 

\begin{figure}[th]
  \centering
  \includegraphics[width=0.95\textwidth]{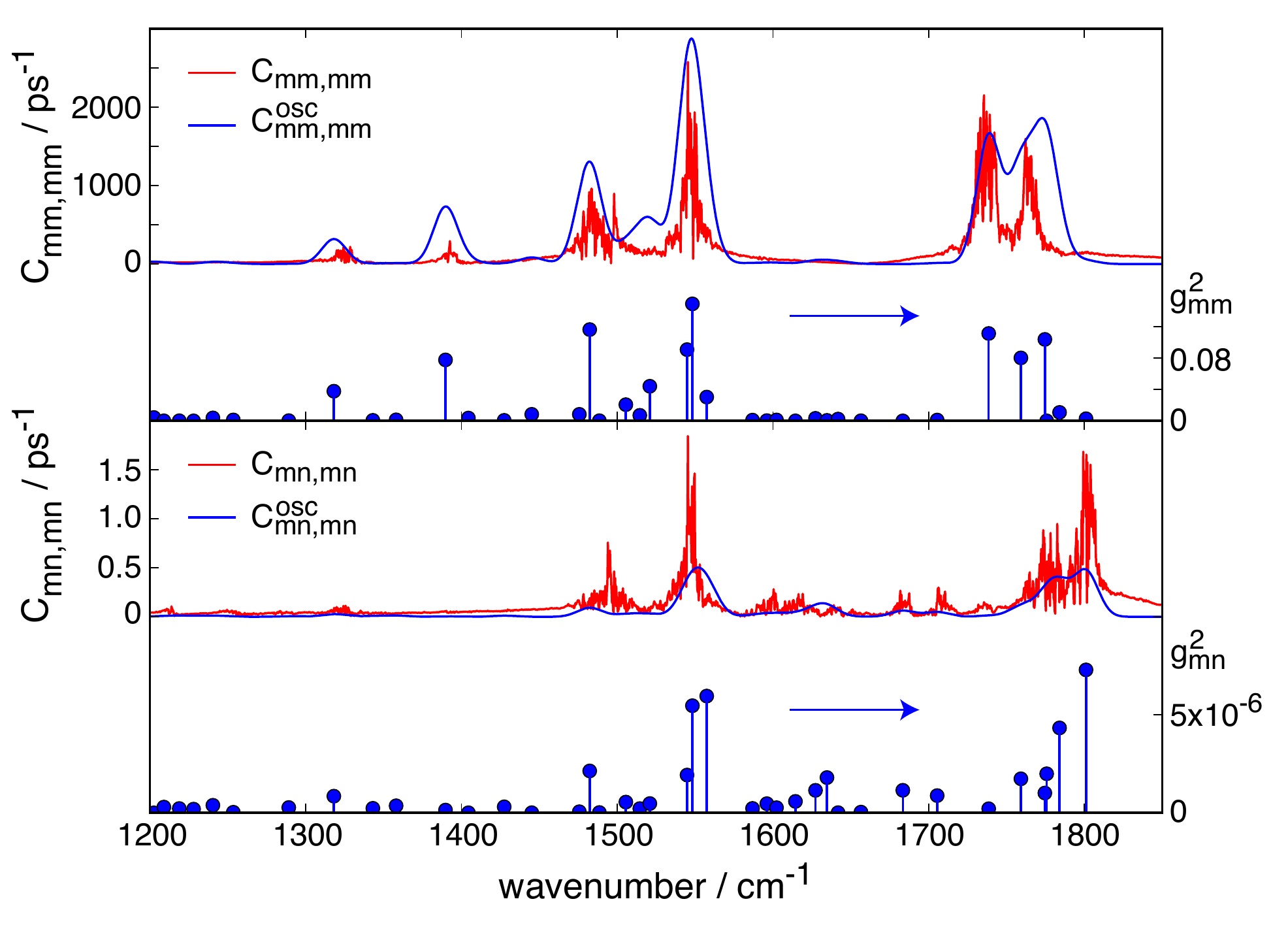}
  \caption{Comparison of DFTB trajectory based spectral distributions, Eq.~\eqref{eq:spec_dens}, with effective harmonic oscillator model, Eq.~\eqref{eq:spec_dens_HO}, for the step-like motif. For better comparison, $C^{\rm osc}_{mm,mm}$ has been broadened by a Gaussian window with 144~\cm{} variance. Also shown are the distribution of couplings, which enter the SD, Eq.~\eqref{eq:SD}. For results on the other motifs, see Supplementary Material.
}
  \label{fig:HOmapping}
\end{figure}

Within dissipative quantum dynamics and spectroscopy it is customary to use SDs, ${\mathcal J}_{kl,mn}(\omega)$, for parametrization of a system-bath Hamiltionian.~\cite{may11} In order to obtain SDs we use the dynamically defined generalized normal modes~\cite{mathias11_2028,plotz17_084112} to calculate  ${\mathcal J}_{mn,mn}(\omega)$  according to Eq.~\eqref{eq:gmn}. The thus obtained SDs (i.e. distributions of squared coupling strengths) are plotted in Fig.~\ref{fig:HOmapping} for the on- and inter-site  cases.  This figure also contains plots of $C_{mn,mn}(\omega)$ and $C_{mn,mn}^{\rm osc}(\omega)$, the latter has been uniformly broadened for better comparison. Note that the width of the peaks in $C_{mn,mn}(\omega)$ is determined by both, the inhomogeneous broadening and the coupling between intra- and intermolecular modes. In principle both effects could be modeled by a multi-mode Brownian oscillator ansatz for the SD.~\cite{mukamel95_} Since the small size of the simulation box doesn't give a good representation for the phonon density of states, this ansatz is not followed here. Overall, there is a fairly good agreement between $C_{mn,mn}(\omega)$ and $C_{mn,mn}^{\rm osc}(\omega)$ as far as the position and intensity of the strongest peaks are concerned. Since the assumption of linear coupling between electronic and nuclear DOF is well justified for the present system, it is very well suited for a mapping onto an effective harmonic oscillator bath. In that sense, $C_{mn,mn}^{\rm osc}(\omega)$ could be considered to be even more accurate than $C_{mn,mn} (\omega)$ because it is not based on the classical approximation for the correlation function.

Finally, note the rather different magnitudes of on-and intersite couplings. As a consequence the the actual properties of the FE will be dominated by the on-site SDs.

\begin{table}
\label{tab:modes}
\begin{tabular}{|c|c|c|c|c|}
\hline 
$\tilde{\nu}_{\rm{DFTB}}$ (\cm) & $\tilde{\nu}_{\rm{FF}}$ (\cm) & overlap & $S_{\rm{DFTB}}$ & $S_{\rm{FF}}$ \\ 
\hline 
91 & 1440 & 0.261 & 0.186 & 0.002 \\ 
\hline 
114 & 114 & 0.586 & 0.287 & 0.033 \\ 
\hline 
146 & 163 & 0.586 & 0.200 & 0.033 \\ 
\hline 
164 & 594 & 0.235 & 0.078 & 0.005\\ 
\hline 
208 & 277 & 0.705 & 0.060 &  0.011 \\ 
\hline 
256 & 316 & 0.673 & 0.062 & 0.000 \\ 
\hline 
263 & 269 & 0.801 & 0.023 & 0.000 \\ 
\hline 
338 & 372 & 0.332 & 0.044 & 0.003 \\ 
\hline 
583 & 1262 & 0.239 & 0.040 & 0.000 \\ 
\hline 
1390 & 1537 & 0.390 & 0.040 & 0.002\\ 
\hline 
1482 & 1544 & 0.548 & 0.058 & 0.005 \\ 
\hline 
1521 & 1781 & 0.208 & 0.022 & 0.000 \\ 
\hline 
1545 & 1621 & 0.618 & 0.045 & 0.000 \\ 
\hline 
1548 & 1805 & 0.400 & 0.075 & 0.007 \\ 
\hline 
1739 & 2007 & 0.689 & 0.056 & 0.000 \\ 
\hline 
1759 & 1765 & 0.586 & 0.040 & 0.000 \\ 
\hline 
1775 & 1782 & 0.429 & 0.052 & 0.000 \\ 
\hline 
\end{tabular} 
\caption{Comparison of DFTB and FF frequencies and Huang-Rhys factors for those modes having a DFTB Huang-Rhys factor larger than 0.02. The correspondence is established based on the Kuhn-Munkres algorithm, cf. Fig.~\ref{fig:overlap}.}
\end{table}

\begin{figure}[th]
  \centering
  \includegraphics[width=0.7\textwidth]{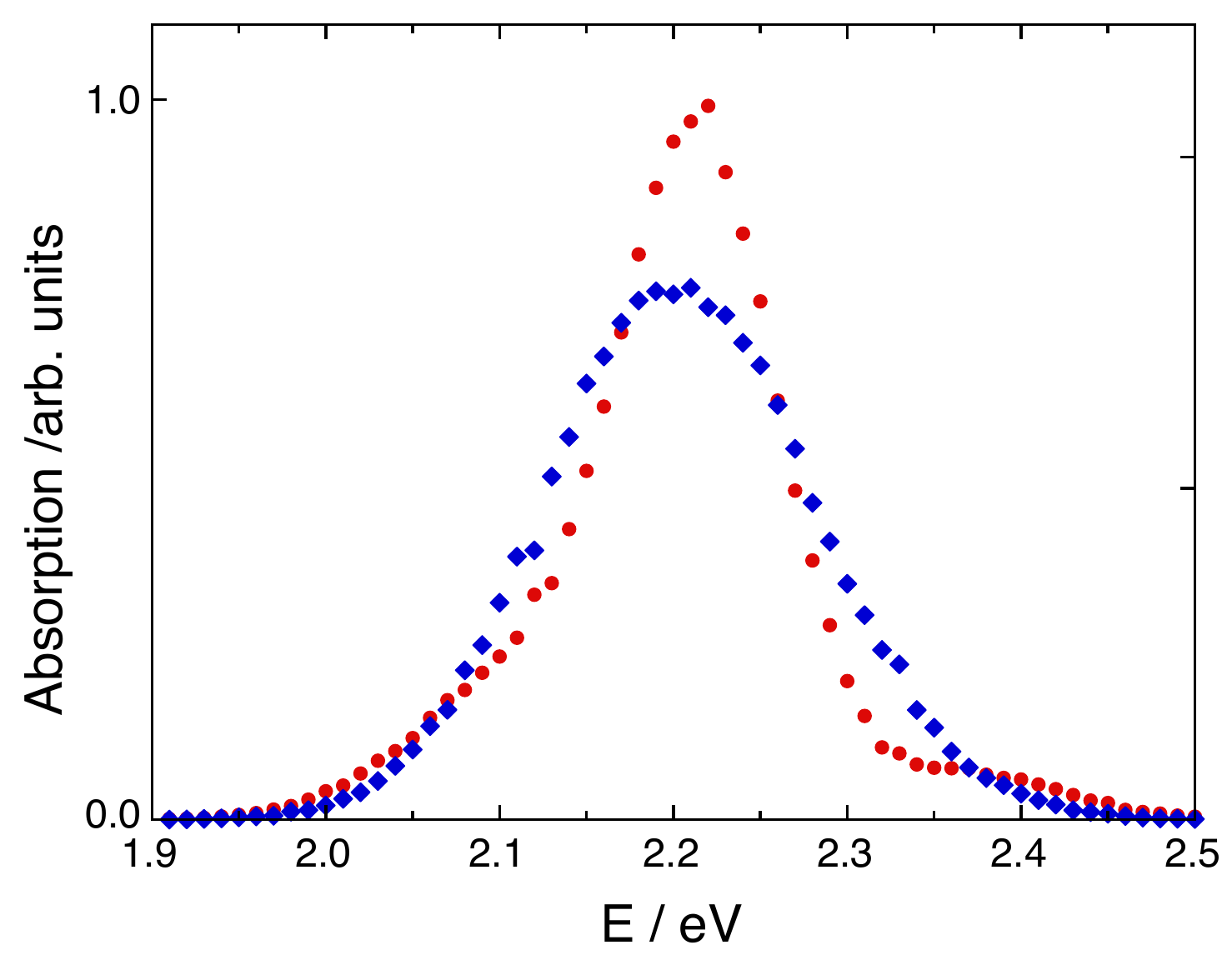}
  \caption{Absorption spectrum according to  the distribution of eigenenergies and transition dipole matrix elements in the simulation box using 10000 samples from the DFTB trajectory (red dots: TBFE spectrum, blue squares: no CC).}
  \label{fig:spectrum}
\end{figure}

\subsection{Absorption Spectrum}
In order to provide a comparison of the present model with experiment, we have calculated the absorption spectrum based on sampling the eigenenergies and transition dipole matrix elements of the TBFE Hamiltonian along the DFTB trajectory. The resulting spectrum is shown in Fig.~\ref{fig:spectrum} (red dots) together with a spectrum obtained by setting the CC equal to zero (blue squares). Comparing the cases with and without inclusion of the CC we notice that the latter causes a narrowing of the spectrum as well as the appearance of a low-intensity peak at the high-energy edge. 

The experimental absorption spectrum reported in Ref.~\citenum{megow15_5747} for a 12~nm film deposited on an Al$_2$O$_2$ surface differs from the present calculation. It shows two peaks around 2.15 eV and 2.45 eV of approximately the same height, with the latter being substantially broader than the former. Furthermore, there is a long tail towards higher energies. There have been different interpretations of this double peak structure. On the one hand, in Ref.~\citenum{megow15_5747} it was shown that a model including heterogeneity of different micro-crystallites is able to reproduce the double peak structure. 
On the other hand, there are a number of publications presenting spectra for crystals of perylene derivatives which demonstrate that a strong coupling between FE and CT states
(see e.g. Ref.~\citenum{hoffmann02_024305,gisslen09_115309}) would also result in a double peak structure.
Here, we cannot resolve this issue, since the correct computation of FE-CT mixing in molecular crystals would require the computation of environmentally induced inductive and dispersive effects on both, FE and CT states. The present model merely provides a reference for the line-broadening effect due to thermal fluctuations in case of a pure FE model.

\section{Conclusions}
%
\label{sec:conclusions}
An all-DFTB based method has been introduced, which allows for the parametrization of the Frenkel exciton Hamiltonian describing excitation energy transfer in molecular assemblies. It includes the influence of thermal fluctuations of the nuclear degrees of freedom. Compared with mixed approaches that treat nuclear dynamics and electronic excitations on different footing, the present protocol provides a consistent framework that is systematically improvable. 

The new protocol has been applied to a model of a PTCDI crystal. For the purpose of illustration we have chosen a rather small simulation box, such that certain aspects of the real PTCDI system are not covered (e.g. coupling to phonon modes). Further, after a detailed discussion of the potential role of CT excitation, we  restricted ourselves to Frenkel excitons only. Comparing FF and DFTB simulations the pitfalls of mixed simulations  could be illustrated in terms of differences in intramolecular modes frequencies and associated Huang-Rhys factors. This is reflected in the spectral distributions of excitation energies and CCs, which are key for describing spectroscopy and transport phenomena. 

SDs are central to the description of the coupled dynamics of electronic excitations and nuclear degrees of freedom.~\cite{may11} They are commonly determined by invoking the classical limit for the nuclei for the calculation of electronic gap correlation functions. In contrast to their quantum counterparts, classical correlation functions are  real-valued. A widespread strategy to correct for quantum effects is to assume that the classical correlation functions are equal to  the real part of the quantum correlation function.~\cite{may11} If applied to SDs, this procedure may be problematic as pointed out by Coker and co-workers.~\cite{rivera13_5510} As an alternative, one may attempt to include quantum effects into the classical trajectory simulation of the gap-correlation function using path integral methods as discussed, for instance, in Ref.~\citenum{karsten18_102337}. In the present contribution we have demonstrated another strategy which builds on the direct determination of the SD without resorting to correlation functions. It is based on the introduction of dynamically-defined normal modes using the approach of Mathias and Baer.~\cite{mathias11_2028} This maps the dynamics onto an effective harmonic oscillator bath, whose correlation functions are known analytically. Of course, these modes are still constructed using classical molecular dynamics. Whether quantum effects in the sampling of the vibrational distribution have a significant influence on the character of the displacement vectors remains to be shown. Here, the recently suggested use of the linearized semiclassical initial-value representation for sampling the initial quantum distribution  in SD simulations~\cite{gottwald18_} could provide a starting point.
%
\section*{Acknowledgements}
Financial support by the Deutsche Forschungsgemeinschaft through projects ME 4215/2-3 (J.M.) and Sfb 652 (P.-A.P., O.K.) is gratefully acknowledged.	
%
\begin{suppinfo}

The following files are available free of charge.
\begin{itemize}
  \item PTCDI2${}\_$SM.pdf: Contains results for spectral distributions like in Fig. 6 for all considered structural motifs.
\end{itemize}

\end{suppinfo}

\providecommand{\latin}[1]{#1}
\makeatletter
\providecommand{\doi}
  {\begingroup\let\do\@makeother\dospecials
  \catcode`\{=1 \catcode`\}=2 \doi@aux}
\providecommand{\doi@aux}[1]{\endgroup\texttt{#1}}
\makeatother
\providecommand*\mcitethebibliography{\thebibliography}
\csname @ifundefined\endcsname{endmcitethebibliography}
  {\let\endmcitethebibliography\endthebibliography}{}

%

%

\end{document}